\documentclass[11pt,a4paper]{article}
\usepackage{amssymb} \usepackage{amsmath} \usepackage{graphicx}
\usepackage{epsfig,latexsym}
\begin{document}

\def\bef{\begin{figure}}
\def\eef{\end{figure}}
\newcommand{\ans}{ansatz }
\newcommand{\be}[1]{\begin{equation}\label{#1}}
\newcommand{\beq}{\begin{equation}}
\newcommand{\ee}{\end{equation}}
\newcommand{\beqn}[1]{\begin{eqnarray}\label{#1}}
\newcommand{\eeqn}{\end{eqnarray}}
\newcommand{\bd}{\begin{displaymath}}
\newcommand{\ed}{\end{displaymath}}
\newcommand{\mat}[4]{\left(\begin{array}{cc}{#1}&{#2}\\{#3}&{#4}
\end{array}\right)}
\newcommand{\matr}[9]{\left(\begin{array}{ccc}{#1}&{#2}&{#3}\\
{#4}&{#5}&{#6}\\{#7}&{#8}&{#9}\end{array}\right)}
\newcommand{\matrr}[6]{\left(\begin{array}{cc}{#1}&{#2}\\
{#3}&{#4}\\{#5}&{#6}\end{array}\right)}
\newcommand{\cvb}[3]{#1^{#2}_{#3}}
\def\lsim{\raise0.3ex\hbox{$\;<$\kern-0.75em\raise-1.1ex
e\hbox{$\sim\;$}}}
\def\gsim{\raise0.3ex\hbox{$\;>$\kern-0.75em\raise-1.1ex
\hbox{$\sim\;$}}}
\def\abs#1{\left| #1\right|}
\def\simlt{\mathrel{\lower2.5pt\vbox{\lineskip=0pt\baselineskip=0pt
           \hbox{$<$}\hbox{$\sim$}}}}
\def\simgt{\mathrel{\lower2.5pt\vbox{\lineskip=0pt\baselineskip=0pt
           \hbox{$>$}\hbox{$\sim$}}}}
\def\unity{{\hbox{1\kern-.8mm l}}}
\newcommand{\eps}{\varepsilon}
\def\ep{\epsilon}
\def\ga{\gamma}
\def\Ga{\Gamma}
\def\om{\omega}
\def\omp{{\omega^\prime}}
\def\Om{\Omega}
\def\la{\lambda}
\def\La{\Lambda}
\def\al{\alpha}
\newcommand{\ov}{\overline}
\renewcommand{\to}{\rightarrow}
\renewcommand{\vec}[1]{\mathbf{#1}}
\newcommand{\vect}[1]{\mbox{\boldmath$#1$}}
\def\tm{{\widetilde{m}}}
\def\mcirc{{\stackrel{o}{m}}}
\newcommand{\Dm}{\Delta m}
\newcommand{\dm}{\varepsilon}
\newcommand{\tanb}{\tan\beta}
\newcommand{\nbar}{\tilde{n}}
\newcommand\PM[1]{\begin{pmatrix}#1\end{pmatrix}}
\newcommand{\up}{\uparrow}
\newcommand{\down}{\downarrow}
\def\omE{\omega_{\rm Ter}}
%

\newcommand{\Dsusy}{{susy \hspace{-9.4pt} \slash}\;}
\newcommand{\DCP}{{CP \hspace{-7.4pt} \slash}\;}
\newcommand{\mc}{\mathcal}
\newcommand{\gr}{\mathbf}
\renewcommand{\to}{\rightarrow}
\newcommand{\gtc}{\mathfrak}
\newcommand{\wh}{\widehat}
\newcommand{\br}{\langle}
\newcommand{\kt}{\rangle}


\def\lsim{\mathrel{\mathop  {\hbox{\lower0.5ex\hbox{$\sim$}
\kern-0.8em\lower-0.7ex\hbox{$<$}}}}}
\def\gsim{\mathrel{\mathop  {\hbox{\lower0.5ex\hbox{$\sim$}
\kern-0.8em\lower-0.7ex\hbox{$>$}}}}}

\def\nn{\\  \nonumber}
\def\de{\partial}
\def\brf{{\mathbf f}}
\def\bbf{\bar{\bf f}}
\def\bF{{\bf F}}
\def\bbF{\bar{\bf F}}
\def\bA{{\mathbf A}}
\def\bB{{\mathbf B}}
\def\bG{{\mathbf G}}
\def\bI{{\mathbf I}}
\def\bM{{\mathbf M}}
\def\bY{{\mathbf Y}}
\def\bX{{\mathbf X}}
\def\bS{{\mathbf S}}
\def\bb{{\mathbf b}}
\def\bh{{\mathbf h}}
\def\bg{{\mathbf g}}
\def\bla{{\mathbf \la}}
\def\bmu{\mathbf m }
\def\by{{\mathbf y}}
\def\bmu{\mbox{\boldmath $\mu$} }
\def\bsig{\mbox{\boldmath $\sigma$} }
\def\bunity{{\mathbf 1}}
\def\cA{{\cal A}}
\def\cB{{\cal B}}
\def\cC{{\cal C}}
\def\cD{{\cal D}}
\def\cF{{\cal F}}
\def\cG{{\cal G}}
\def\cH{{\cal H}}
\def\cI{{\cal I}}
\def\cL{{\cal L}}
\def\cN{{\cal N}}
\def\cM{{\cal M}}
\def\cO{{\cal O}}
\def\cR{{\cal R}}
\def\cS{{\cal S}}
\def\cT{{\cal T}}
\def\eV{{\rm eV}}
%





\large
 \begin{center}
 {\Large \bf More about the Instanton/Soliton/Kink correspondence }
 \end{center}

 \vspace{0.1cm}

 \vspace{0.1cm}
 \begin{center}
{\large Andrea Addazi}\footnote{E-mail: \,  andrea.addazi@infn.lngs.it} \\
{\it \it Dipartimento di Fisica,
 Universit\`a di L'Aquila, 67010 Coppito, AQ \\
LNGS, Laboratori Nazionali del Gran Sasso, 67010 Assergi AQ, Italy}
\end{center}

\vspace{1cm}
\begin{abstract}
\large

We demonstrate that all gauge instantons in a $d=3+1$ Yang-Mills theory, 
with generic topological vacuum  charge K, correspond to soliton solutions
and kink scalar fields 
in  $d=4+1$ space-time.

\end{abstract}

\baselineskip = 20pt

\section{Introduction}

G.~Dvali, H.~B.~Nielsen and N.~Tetradis (DNT)
have studied the dynamical localization 
of a $d=3$ gauge field in a lower dimensional 
2-d surface \cite{Dvali:2007nm}. 
They have considered the same set-up suggested by 
  G.~R.~Dvali and M.~A.~Shifman \cite{Dvali:1996xe}:
 $SU(2)$ YM theory Higgsed down to a $U(1)$
 inside
  a region $\Sigma$ closed by two 2-d domain walls. 
  Inside $\Sigma$ the Higgsation corresponds to the presence 
  of a kink solution charged with respect to $SU(2)$ but leaving 
  massless a $U(1)$ gauge subgroup. 
The kink solution corresponds to the presence of a 
t'Hooft-Polyakov monopole in the bulk region $\Sigma$. 
The kink/soliton correspondence was well known 
in literature of domain walls, topological solutions 
and D-branes 
\cite{Hanany:2003hp,Tong:2005un,Sen:1998sm}.
Finally, DNT have shown a correspondence 
among the t'Hooft-Polyakov monopole \cite{TP1,Polyakov:1974ek} in the bulk and 
a gauge instanton solution in the lower dimensional theory 
localized on the domain walls.
The gauge instanton solution describes the quantum tunneling process 
of the monopole outside the bulk region $\Sigma$.
Of course, these arguments can work only considering  that domain walls are at a distance 
$l>>\rho$, where $\rho$ is the effective size of the monopole solution. 
In this regime, the problem is in semiclassical approximation regime and 
 one can neglect quantum non-perturbative corrections to the system. 
The tunneling probability is expected to be exponentially 
suppressed as in standard WKB semiclassical approximation.

In this paper, we will suggest a general proof of the instanton/soliton correspondence 
to a generic $d=4$ YM theory. We will demonstrate that all instanton solutions in 
a $d=4$ YM theory are in $1\leftrightarrow 1$ correspondence with 
$d=5$ systems composed of two $d=4$ domain walls closing 
a solitonic solutions inside the bulk $\Sigma$. 
In our proof, we will use geometric methods 
developed by   M.F. Atiyah, V.G. Drinfeld, N.J. Hitchin, Y.I. Manin,
known in literature as ADHM construction \cite{ADHM1,ADHM2}. 
The ADHM construction is a general formal method to construct 
all gauge instantons of a generic YM gauge theory.
In particular, ADHM construction uses twistor/quaternionic formalism 
in order to classify all possible  fiber/connection gauge fields on the 
Minkonski space-time. 
Our strategy will be the following: 
using the kink/soliton correspondence, 
we will demonstrate that the unitary matrix $\mathcal{U}_{A}$,
diagonalizing the tachyon condensate matrix $|\mathcal{T}|^{2}$,
will have exactly the same moduli of the ADHM ansatz matrix 
($A_{\mu}=\frac{1}{g}\mathcal{U}_{A}\partial_{\mu}\mathcal{U}_{A}$,
where $g$ is the YM gauge coupling),
under quite general hypothesis. 
In other words, the number of moduli parametrizing the shape 
of the tachyon kink is in $1\leftrightarrow 1$ correspondence 
with moduli parametrizing the instantonic solution. 
This correspondence among kinks and instantons 
will be found for a generic instanton 
with generic topological charge $K$. 
In particular, an instanton with $K=1$ is recovered 
from a kink charged with respect to $SU(N+2)\times U(2)$, 
while for generic $K$ from a kink charged with respect to $SU(N+2K)\times U(2K)$. 

\section{ADHM construction: a short review}

ADHM construction is a powerful geometric method allowing 
to construct all gauge instantons. 
It uses $x^{\mu}$ in Euclidean space-time coordinates
in quaternionic notation:
\be{quaternionic}
x_{ij}^{\mu}= \left( \begin{array}{cc} z_{2}
 & z_{1}
\ \\ -\bar{z}_{1} & \bar{z}_{2} \ \\
\end{array} \right)
\ee
Let us consider a $SU(N)\times U(K)$ YM theory in $d=4$.
The ADHM construction is based on the 
following ansatz: 
\be{ansatz}
A_{\mu}(x)=\frac{1}{g}\bar{\mathcal{U}}_{A}\partial_{\mu}\mathcal{U}_{A}
\ee
where $\mathcal{U}_{A}$ is a $(N+2K) \times 2K$ complex matrix.
Columns of $\mathcal{U}_{A}$ are the orthonormal 
vectors, basis of the N-dimensional null space of the Dirac operator
$\Delta$:
\be{Basis}
\Delta \mathcal{U}_{A}=0
\ee
where $\Delta$ is a $2K \times (N+2K)$ matrix. 
It turns out that $\Delta$ is just a linear matrix function of 
$x_{ij}^{\mu}$, i.e. it can be written in terms of matrices $I,J, B_{1,2}$ and quaternionic variables:
\be{Deltaaa}
\Delta= \left( \begin{array}{ccc} I & B_{2}+z_{2} & B_{1}+z_{1}
\ \\ J^{\dagger} & -B_{1}^{\dagger}-\bar{z}_{1} & B_{2}^{\dagger}+\bar{z}_{2} \ \\
\end{array} \right)
\ee
with ADHM constraints 
\be{c1}
0=\mu_{R}=2\mu_{3}=I^{\dagger}I-J^{\dagger}J+[B_{1},B_{1}^{\dagger}]+[B_{1},B_{1}^{\dagger}]
\ee
\be{c2}
0=\mu_{C}=\mu_{1}+i\mu_{2}=IJ+[B_{1},B_{2}]
\ee
Eqs.(\ref{c1})-(\ref{c2}) reduce the redundancy of parameters exactly to the moduli space
of gauge instantons. 
Eqs.(\ref{c1})-(\ref{c2}) correspond to the factorization 
\be{DD}
\Delta \Delta^{\dagger}={\rm diag}\{F(x) , F(x)\}
\ee
where $F(x)$ is a $k\times k$ Hermitian matrix. 

So that, the ADHM construction uses the following ADHM data: 
complex matrices $\mathcal{U}_{A}$ composed of the orthonormal vectors of the null space
of the Dirac operator $\Delta$ ;
$B_{1,2}$ which are $K\times K$ complex matrices;
$I,J$ $K\times N$ and $N\times K$ complex matrices; 
constraints $\mu_{R,C}=0$. 
The metric of the instanton moduli space is contained 
in the flat metric of $B,I,J$.

\section{Main argument}

Let us consider a $SU(N+2K)\times SU(2K)$ YM theory.
We suppose inside the bulk region $\Sigma$ 
a tachyonic kink potential connecting to 
asymptotic vacua $\pm v$ outside $\Sigma$.
The tachyonic kink has a generic sombrero
potential $V(\mathcal{T})$, 
where $\mathcal{T}$ is the scalar field 
with a ground kink solution in $\Sigma$. 
The Tachyon field is chosen as an adjoint 
representation of $SU(N+2K)\times SU(2K)$,
i.e. it is a $(N+2K)\times  2K$ matrix. 
This implies that the combination 
$\mathcal{T}\mathcal{T}^{\dagger}$ 
is a $(N+2K)\times (N+2K)$ matrix 
with at least $2K$ zero eigenvalues, 
while
$N+2K$ eigenvalues are exactly equal to 
the one of $\mathcal{T}^{\dagger}\mathcal{T}$. 
So that, $\mathcal{T}\mathcal{T}^{\dagger}$
can be diagonalize in irriducible blocks by a unitary rotation matrix: 
\be{UTTU}
\mathcal{U}_{A}\mathcal{T} \mathcal{T}^{\dagger}\mathcal{U}_{A}^{\dagger}= \left( \begin{array}{cc} 0
 & 0
\ \\ 0 & \mathcal{T}^{\dagger}\mathcal{T} \ \\
\end{array} \right) \end{equation}
We assume as a hypothesis that $\mathcal{T}^{\dagger}\mathcal{T}$ is definite positive. 
This implies that all eigenvalues of $\mathcal{T}^{\dagger}\mathcal{T}$ 
cannot be negative or null values, i.e. $\lambda_{i}> 0$. 
Now let us define the matrix $\Lambda=\sqrt{\mathcal{T}^{\dagger}\mathcal{T}}$.
From proprieties of $\mathcal{T}^{\dagger}\mathcal{T}$,
$\Lambda$ is hermitian  with a $2^{2K}$ degeneracy of its eigenvalues 
$\pm \sqrt{\lambda_{l}}$. 
Let us consider only positive eigenvalues of $\Lambda$, 
up to discrete symmetries $Z_{2}^{2K}$ subgroup of $U(2K)$. 
Under these assumptions, $\mathcal{U}_{A}$ can be decomposed 
as eigenvectors of the null space $\mathcal{N}$ of $\mathcal{T}^{\dagger}$ operators
an the hortogonal to the null space $\mathcal{N}^{T}$:
\be{U1}
\mathcal{U}_{A}=(\mathcal{V},\mathcal{W})
\end{equation}
\be{V1}
\mathcal{T}^{\dagger}\mathcal{V}=0
\end{equation}
\be{W}
\mathcal{W}=\mathcal{T}\Lambda^{-1}
\end{equation}
where $\Lambda^{-1}$ can be defined under definite positiveness 
assumption assumed above.
Now, let us impose the unitarity condition on the $\mathcal{U}_{A}$ matrix:
\be{Uncond}
{\bf I}_{N+2K}=\mathcal{U}_{A}^{\dagger}\mathcal{U}_{A}=
\left( \begin{array}{cc} \mathcal{V}^{\dagger}\mathcal{V}
 & \mathcal{V}^{\dagger}\mathcal{T}\Lambda^{-1}
\ \\ \Lambda^{-1}\mathcal{T}^{\dagger}\mathcal{V} & \Lambda^{-1}\mathcal{T}^{\dagger}\mathcal{T}\Lambda^{-1} \ \\
\end{array} \right)
\ee
\be{Uncond2}
{\bf I}_{2K}=\mathcal{U}_{A}\mathcal{U}_{A}^{\dagger}=\mathcal{V}\mathcal{V}^{\dagger}+\mathcal{T}\Lambda^{-2}\mathcal{T}^{\dagger}
\ee
where ${\bf I}_{N+2K}, {\bf I}_{2K}$ are the $(N+2K)\times (N+2K)$ identity matrices. 
We will demonstrate that Eqs.(\ref{Uncond})-(\ref{Uncond2})
are nothing but ADHM equations of a generic K-instanton of a 
$SU(N)$ gauge theory.  In other words, $\mathcal{U}_{A}$ corresponds to the matrix of ortho-normal eigenvectors 
of the Dirac operators $\Delta$. So that, from $\mathcal{U}_{A}$ we will construct the fiber gauge connection 
field $A_{\mu}$, according to ADHM ansatz. 

The Eqs. (\ref{Uncond})-(\ref{Uncond2}) guarantee that the matrix $\mathcal{T}$ can be decomposed in the basis of $\sigma^{\mu}$
as 
\be{Uncond3}
\mathcal{T}(x)=
c\left( \begin{array}{c} \alpha
\ \\ (\beta^{\mu}-x^{\mu})\sigma^{\mu} \ \\
\end{array} \right)
\ee
It will turn out that $\alpha, \beta^{\mu}$ are the same 
instantonic data of the ADHM construction:
\be{Mkn}
\mathcal{M}_{N,K}=\{\alpha, \beta^{\mu} \}/U(k)
\ee
In order to show the correspondence of parameters 
between (\ref{Uncond3}) and (\ref{Mkn}), 
let us expand the combination $\mathcal{T}^{\dagger}\mathcal{T}$
in basis of Pauli matrices as
\be{expandP}
\mathcal{T}^{\dagger}\mathcal{T}=c^{2}\mu_{0}\times {\bf I}_{2}+c^{2}\sum_{i=1}^{3}\mu_{i}\times \tau_{i}
\ee
where $\mu_{0},\mu_{i}$ are $K\times K$ matrices and $c$ is a constant. 
Now, from the 
general proprieties of decompositions in Pauli basis, 
we can rewrite coefficient matrices in terms of new matrices $I,J,B_{n}$ and quaternionic variables $z,\bar{z}$:
\be{mur}
\mu_{R}=2\mu_{3}=I^{\dagger}I-J^{\dagger}J+[B_{1},B_{1}^{\dagger}]+[B_{2},B_{2}^{\dagger}]
\ee
\be{muC}
\mu_{C}=\mu_{1}+i\mu_{2}=IJ+[B_{1},B_{2}]
\ee
\be{mu0}
\mu_{0}=\frac{1}{2}\left\{II^{\dagger}+JJ^{\dagger}+\frac{1}{2}\sum_{n=1}^{2}\{B_{n}-z_{a},B_{n}^{\dagger}-z_{a}^{\dagger} \} \right\}
\ee
Bu these conditions are nothing but the ADHM equations. 
In ADHM construction, all gauge instantons are constructed 
using the constrains 
\be{muzero}
\mu_{C}=\mu_{R}=0
\ee
while $\mathcal{T}$ is Linearly dependent to the $\Delta$ operator as 
\be{TDelta}
\mathcal{T}=c\Delta
\ee
where $\Delta$ is the Dirac operator of ADHM. 
In particular 
\be{Uncond4}
\mathcal{T}(x)=
c\left( \begin{array}{ccc} I & B_{2}+z_{2} & B_{1}+z_{1}
\ \\ J^{\dagger} & -B_{1}^{\dagger}-\bar{z}_{1} & B_{2}^{\dagger}+\bar{z}_{2} \ \\
\end{array} \right)
\ee
that is  equivalent to (\ref{Uncond3}) with 
\be{equivalence}
B_{1}=\beta^{2}+i\beta^{1},\,\,\,\,B_{2}=\beta^{0}+i\beta^{3}
\ee
\be{Z12}
z_{1}=x^{2}+ix^{1},\,\,\,\,z_{2}=x^{0}+ix^{3}
\ee

\subsection{t'Hooft-Polyakov monopole and BPST instantons}

Let us consider a tachyon kink with a center distribution 
\be{tachyonkinkBPST}
c\left( \begin{array}{ccc} \rho \times {\bf I}_{2} 
\ \\ (\alpha^{\mu}-x^{\mu})\times \sigma_{\mu} \ \\
\end{array} \right)
\ee
with $\mu_{r,c}=0$ and 
$\mu_{0}=|\alpha-x|^{2}+\rho^{2}$. 
Such a tachyon profile correspond to 
a t'Hooft-Polyakov monopole confined 
into the region $\Sigma$ closed by two domain walls. 
According to the construction given above, 
we obtain the corresponding unitary matrix 
\be{UAfoll}
\mathcal{U}_{A}=\frac{1}{\sqrt{\mu_{0}}}\left( \begin{array}{cc} (\alpha^{\mu}-x^{\mu})\sigma_{\mu} & \rho \times {\bf I}_{2}
\ \\ -\rho \times {\bf I}_{2} & (\alpha^{\mu}-x^{\mu})\sigma_{\mu}
 \ \\
\end{array} \right)
\ee
But this is nothing but the standard BPST solution for $N=2$ and $K=1$
\cite{BPST}
In particular the corresponding field is 
\be{AmuBPST}
A_{\mu}^{a}=\frac{2}{g}\frac{\eta_{\mu\nu}^{a}(x-z)_{\nu}}{(x-z)^{2}+\rho^{2}}
\ee
In particular the instantonic field strenght has the form 
\be{Fmunu}
\mathcal{F}_{\mu\nu}=\frac{2i\rho^{2}\eta_{\mu\nu}^{i}\tau_{i}}{g(|x-\alpha|^{2}+\rho^{2})^{2}}
\ee
Note that the number of moduli for the 5d monopole corresponds to the 4 spontaneously 
broken generator of spatial translation plus 
the only rotation in the internal $SU(2)$ which leaves
the condensate intact, i.e. 4+1 moduli. 
On the other hand, Moduli of 4d BPST solution 
are $4+1$: $x_{0}^{\mu},\rho$, corresponding to the center and the radius of the instanton solution.

\section{Conclusions and outlooks}

In this paper, 
we have demonstrated a correspondence 
among gauge instantons in 4-d YM theories 
and soliton in higher dimensional theories.
The soliton corresponds to a kink solution 
localized in a region $\Sigma$ closed by two 
domain walls. 
We have shown that the number of moduli describing 
the kink solution is exactly coincident to the 
number of moduli describing the position and 
shape of the gauge instanton. 
As suggested by G.~Dvali, H.~B.~Nielsen and N.~Tetradis (DNT), 
the instanton solution describes the quantum tunneling process 
of a soliton outside the domain walls box. 
In our paper, we have shown a formal generalization of their argument 
to more general cases than $SU(2)$ theories studied by DNT, 
by virtue of the ADHM classification of all $SU(N)$ gauge instantons.

Let us comment that
in string theory, there are interesting counter-examples of instantons 
that cannot be obtained by ADHM construction:
Exotic stringy instantons can be constructed as Eucliden D-branes
or E-branes wrapping different n-cycles, on the internal Calabi-Yau compactifications,
then ordinary D-branes
 \cite{Bianchi:2009ij}.
For example, in IIA string theory, Exotic stringy instantons correspond 
to solitonic $E2$-branes wrapping 3-cycles in the $CY_{3}$. 
The number of cycles is determined by the worldvolume 
action of the $E2$-brane: other numbers of cycles would develop tachyonic instabilities. 
So that, instanton/soliton correspondence could be 
also more general than the one that we have demonstrated in this paper 
in context of gauge theories. The relevance of exotic instantons
in particle physics and baryogenesis was recently discussed 
in our papers 
 \cite{Addazi:2014ila,Addazi:2015ata,Addazi:2015rwa,Addazi:2015hka,Addazi:2015yna,Addazi:2015eca,Addazi:2015fua,Addazi:2015oba,Addazi:2015goa,Addazi:2016xuh,Addazi:2016mtn}: we have shown calculable examples of new effective operators induced 
by exotic instantons while not allowed at perturbative level.
Finally, the instanton/soliton correspondence could be extended to semiclassical quantum gravity.
Possible connections among YM instantons and Gravitational instantons
were found by J.~J.~Oh, C.~Park and H.~S.~Yang \cite{Oh:2011nv}.
Gravitational instantons were classified by G. W. Gibbons and S. W. Hawking 
\cite{GH1,GH2}. These solutions can be interpreted as wormholes, 
as argued by Strominger \cite{StromingerW}, Hawking \cite{HawkingW1,HawkingW2} and 
in following papers \cite{Other1,Other2,Other3,Other4,Other5},
with perhaps possible implications in ER=EPR conjecture \cite{Maldacena:2013xja,Susskind:2014yaa}
and in entangled cosmology 
  \cite{Basini:2003bt,Capozziello:2013wea,Capozziello:2014nda}.
In other words, a $S_{2}\times S_{2}$ gravitational instanton mediates
a quantum tunneling of a black hole into a white hole solution. 
Gravitational instantons could also mediate 
tunneling processes of antievaporating unstable Nariai black holes. 
Antievaporation of Nariai black holes was
found by Bousso and Hawking in quantum gravity coupled to a dilaton fields \cite{Bousso:1997wi}. 
This phenomena was re-discovered by Nojiri and Odintsov in classical $f(R)$-gravity \cite{Nojiri:2013su, Nojiri:2014jqa},
and then in several different extension of general relativity 
\cite{Sebastiani:2013fsa,Katsuragawa:2014hda,Katsuragawa:2015sjn,Houndjo:2013qna}.
As shown in Ref.\cite{Addazi:2016prb},
Bekenstein-Hawking radiation cannot be 
emitted by antievaporating black holes, 
i.e. antievaporation cannot be countered by Bekenstein-Hawking evaporation. 
Gravitational instanton/soliton correspondence could be 
useful for an understanding of the final fate of antievaporating black holes.  
From the instanton/soliton correspondence, 
Black hole could be reinterpreted 
as a gravitational soliton, spontaneously breaking 
BMS invariance and carrying quantum soft hairs as BMS moduli
(see \cite{Hawking:2016msc,Dvali:2015rea,Ellis:2016atb}
 for many proposal on linking BMS with quantum soft hairs).
 
We conclude that the instanton/soliton correspondence seems to
be a powerful and general aspect of all gauge theories and 
their extensions (as string theory). 
In this way, instantonic solutions can be visualized as 
extended solitonic configuration localized in a certain region $\Sigma$.
A reinterpretation of instantons as solitons
can provide useful insights in fundamental particle physics and quantum gravity.

\vspace{1cm} 

{\large \bf Acknowledgments} 
\vspace{3mm}

I would like to thank Massimo Bianchi and Gia Dvali
for valuable comments and discussions.
My work was supported in part by the MIUR research
grant "Theoretical Astroparticle Physics" PRIN 2012CPPYP7
and by SdC Progetto speciale
Multiasse La Societ\'a della Conoscenza in Abruzzo PO FSE Abruzzo 2007-2013.


\end{document}